\documentclass[reprint,aps,prx,superscriptaddress,floatfix,showkeys,amsmath,amssymb,amsfonts]{revtex4-2}

\setcitestyle{super}
\usepackage{graphicx}
\usepackage{dcolumn}
\usepackage{bm}
\usepackage[hidelinks]{hyperref}

\begin{document}

\title{Observation of Rashba Magnetism in Ultrathin Ferromagnet-Heavy Metal Bilayers}

\author{Sergei Ivanov}
\email{svivano@emory.edu}
\author{Yiou Zhang}
\author{Guanxiong Chen}
\author{Joshua Peacock}
\affiliation{Department of Physics, Emory University, Atlanta, Georgia 30322, USA}
\author{Vladislav E. Demidov}
\author{Sergej O. Demokritov}
\affiliation{Institute of Applied Physics, University of Münster, 48149 Münster, Germany}
\author{Nicholas Brookes}
\author{Björn Wehinger}
\affiliation{European Synchrotron Radiation Facility, 38043 Grenoble, France}
\author{John William Freeland}
\affiliation{Advanced Photon Source, Argonne National Laboratory, Lemont, Illinois 60439, USA}
\author{Sergei Urazhdin}
\affiliation{Department of Physics, Emory University, Atlanta, Georgia 30322, USA}

\date{\today}

\begin{abstract}
Both magnetism and spin-orbit coupling in systems with broken inversion symmetry lift the spin degeneracy of electronic bands, but the consequences of interplay between these mechanisms remain poorly understood. Here, we show that ultrathin transition ferromagnet-heavy metal bilayers exhibit anomalous temperature- and electric bias-dependent behaviors  in the vicinity of the Curie temperature, inconsistent with the usual Weiss magnetism. Characterization by several complementary techniques and analysis of the dependence on composition reveal that these effects originate from interfacial spin-orbit interaction, which results in the emergence of a state with distinct magnetic and magnetoelectronic properties that can be described as Rashba magnetism. Our findings open a new route for the characterization and control of spin-orbit phenomena in heterostructures enabling the development of efficient spin-orbitronic devices.
\end{abstract}

\keywords{interfaces; Rashba effect; spin-orbit coupling; anomalous Hall effect; spin-galvanic effect}

\maketitle

\section{\label{sec:Introduction}Introduction}

Spin-orbit coupling (SOC) is ubiquitous in condensed matter and plays an important role even in materials based on light elements such as graphene~\cite{PhysRevLett.95.226801}. SOC-driven spin-galvanic effects produce spin-orbit torques (SOTs) on magnetic nanoelements, enabling spin-orbitronic applications in digital and analog magnetic memories~\cite{Kent2015-ec,Ramaswamy2018}, spin wave-based electronics~\cite{Demidov2020}, probabilistic (p-bit) logic~\cite{Shim2017,Camsari2019-va}, neuromorphics~\cite{Chen2020-mq,Grollier2020}, and ultrafast spin lasers~\cite{Dainone2024}. Magnetoelectronic effects mediated by SOC such as anomalous Hall effect (AHE) provide a mechanism for the magnetic state detection. SOTs result from some combination of spin Hall effect (SHE) in SOC materials and Rashba-Edelstein effect (REE) at their interfaces. In SHE, spin current is generated by electric current due to spin-dependent chirality of electron transport~\cite{DYAKONOV1971459,RevModPhys.87.1213}. In REE, spin polarization originates from the chiral spin texture of bands split by the Rashba SOC at interfaces with broken inversion symmetry~\cite{Rashba1984,EDELSTEIN1990233}. Despite different mechanisms, both SHE and REE result in similar SOT effects on magnetic films~\cite{RevModPhys.91.035004}. Likewise, chiral spin-dependent transport mechanisms underlying both SHE and REE result in two contributions to AHE with distinct origins~\cite{RevModPhys.82.1539}.

To achieve efficient spin-orbitronic device operation, it is important to unambiguously identify the two contributions to spin-orbitronic effects and develop approaches to constructively combine them. However, the similarity between their manifestations has aroused a controversy regarding the effects of interfacial SOC. Large Rashba effects were confirmed by direct measurements~\cite{Schulz2019,hu2024chiralquantumrashbasplitting} and observations of efficient spin-charge interconversion without SHE sources~\cite{Haidar2019-zc,PhysRevB.104.184410}. On the other hand, studies of ferromagnet/heavy metal (FM/HM) bilayers suggest that SOC at their interfaces contributes mostly to spin relaxation~\cite{Tao2018,PhysRevLett.122.077201}. An insight into the possible origins of this controversy is provided by the interpretation of the Rashba mechanism as an effective spin-orbit field superimposed in magnetic systems with exchange field~\cite{PhysRevLett.121.136805}. The effects of these fields on spin and orbital dynamics of electrons depend on the relation between their magnitudes. If the exchange field is dominant, the total effective field is only weakly perturbed by the Rashba contribution, resulting in suppression of Rashba effects at magnetic interfaces with large exchange. Similar suppression of spin precession driven by SOC field, the mechanism underlying Dyakonov-Perel spin relaxation, was observed in proximity-magnetized Pt~\cite{PhysRevLett.120.067204}.

To investigate the possibility of such interplay between SOC and magnetism, we study ultrathin-film FM/HM bilayers with Curie temperature suppressed by reduced dimensionality. The effective exchange field is expected to rapidly vary with temperature $T$ in the vicinity of the Curie temperature, enabling control of the relative magnitudes of this field and the effective Rashba field. We report anomalous temperature- and electric bias-dependent magnetic and magnetoelectronic behaviors in the vicinity of the Curie temperature. By analyzing  the dependence on composition and spin fluctuation dynamics, we unambiguously identify these behaviors with interfacial SOC effects. Our observations are consistent with the expected suppression of the Rashba effect by magnetism, suggesting a straightforward approach to characterize the Rashba contribution to spin-galvanic phenomena and its optimization for spin-orbitronic applications.

\begin{figure}
	\includegraphics[width=1\linewidth]{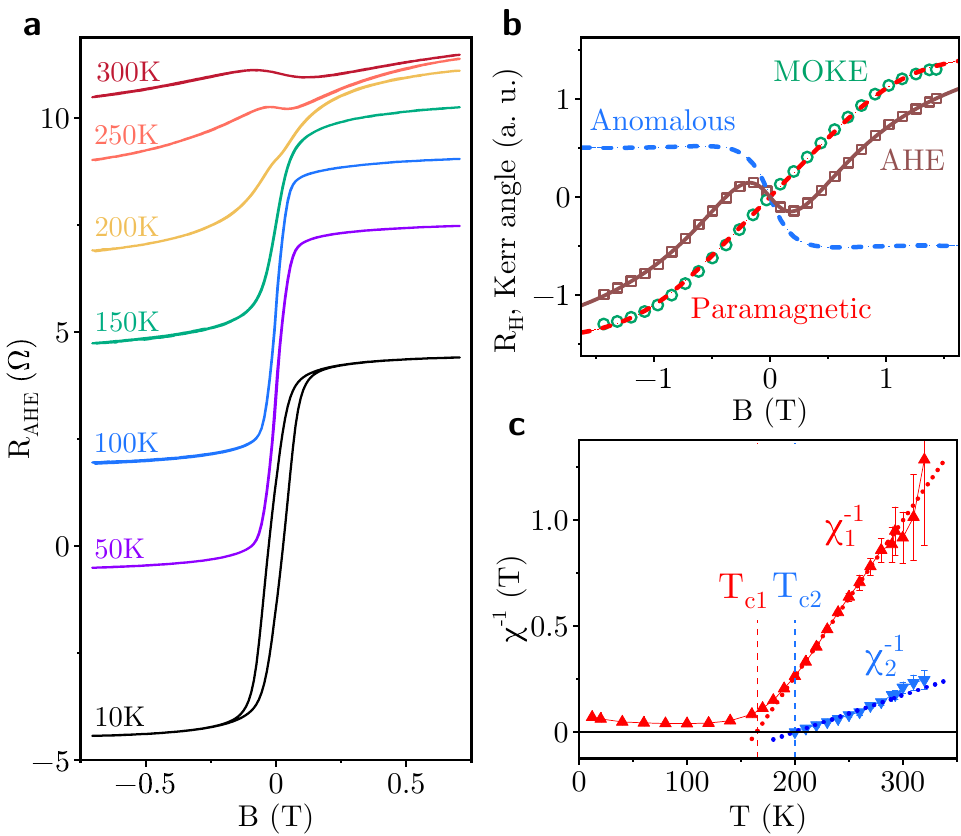}
	\caption{Magnetoelectronic and magneto-optical characterization of a Ti(1.5)Pt(2)CoFeB(0.3)AlO$_x$(2) film. (a) AHE resistance vs field, at the labeled $T$. Curves are offset for clarity. (b) Symbols: AHE resistance (squares) and polar MOKE (circles) vs field, at $T=295$~K. Solid curve: fitting of the AHE data with two sigmoids, dashed lines: sigmoids plotted separately. Linear ordinary Hall background obtained from low temperature high field data is subtracted before fitting. (c) Reciprocal susceptibilities obtained from the AHE data using Eq.~(\ref{eq:ahe_fitting}).
 }\label{fig:AHE}.
\end{figure}

\section{\label{sec:Results}Results}

We studied heterostructures based on ultrathin transition-metal FMs whose thickness was adjusted so that the Curie temperature $T_{c1}$ was close to or below room temperature (RT) $T=295$~K, enabling their straightforward characterization in the vicinity of $T_{c1}$.  Figure~\ref{fig:AHE} shows representative results for a Ti(1.5)Pt(2)CoFeB(0.3)AlO$_x$(2) heterostructure characterized by $T_{c1}=150$~K. Numbers in parentheses are layer thicknesses in nanometers. At $T\ll T_{c1}$, the dependence of AHE resistance $R_{\text{AHE}}$ on field $B$ normal to the film shows a narrow hysteresis consistent with almost compensated magnetic anisotropy [Fig.~\ref{fig:AHE}a], which was verified by X-ray magnetic circular dichroism (XMCD) [see Supplemental Material]. At $T>T_{c1}$, the dependence $R_{\text{AHE}}(B)$ becomes broadened, and a nonlinear feature with the sign of $R_{\text{AHE}}$ opposite to the dominant paramagnetic contribution appears at $T>200$~K at small $B$. It broadens at larger $T$ and becomes increasingly dominant, resulting in a reversed slope of the Hall resistance at modest $B$. Similar anomalous AHE features, whose mechanism remains unexplained, were reported for other heterostructures of magnetic materials with  HMs~\cite{doi:10.1063/1.2827174,PhysRevB.87.104407,PhysRevLett.109.107204,PhysRevB.89.140407,doi:10.1063/1.4916342,PhysRevB.92.165114,10.1007/s12598-022-02166-z}.

The AHE curves are well-approximated by the sum of two sigmoid functions with different widths [Fig.~\ref{fig:AHE}b], indicating two contributions to magnetism with distinct responses to applied field. Magneto-optic Kerr effect (MOKE) curve matches the dominant paramagnetic contribution without an anomalous feature, suggesting that these two contributions originate from different mechanisms. MOKE is dominated by the spin polarization of quasi-localized electrons~\cite{Wolff:13}, while AHE $-$ by the chiral spin-dependent transport properties of the Fermi surface~\cite{RevModPhys.82.1539}. Thus, we tentatively attribute the anomalous feature to magnetism of conduction states, as opposed to the conventional quasi-localized d-electron magnetism in TM FMs~\cite{OHandley1999-ig}. 

We define magnetoelectronic susceptibilities $\chi_1$, $\chi_2$ associated with each contribution as the slopes of the corresponding sigmoid dependences at $B=0$ normalized by their amplitudes [see \textit{Methods}]. The linear temperature dependences of both reciprocal susceptibilities are consistent with the Curie-Weiss law, Fig.~\ref{fig:AHE}c. One intercept is the Curie temperature $T_{c1}$ , and another is $T_{c2}=200$~K describing the anomalous feature. In other structures with different values of $T_{c1}$, $T_{c2}$ remained $30-80$~K higher [see Figs.S1-S3], confirming that the anomalous AHE feature is related to magnetic ordering and is not a merely spin-electronic effect.

\begin{figure}
	\includegraphics[width=0.9\linewidth]{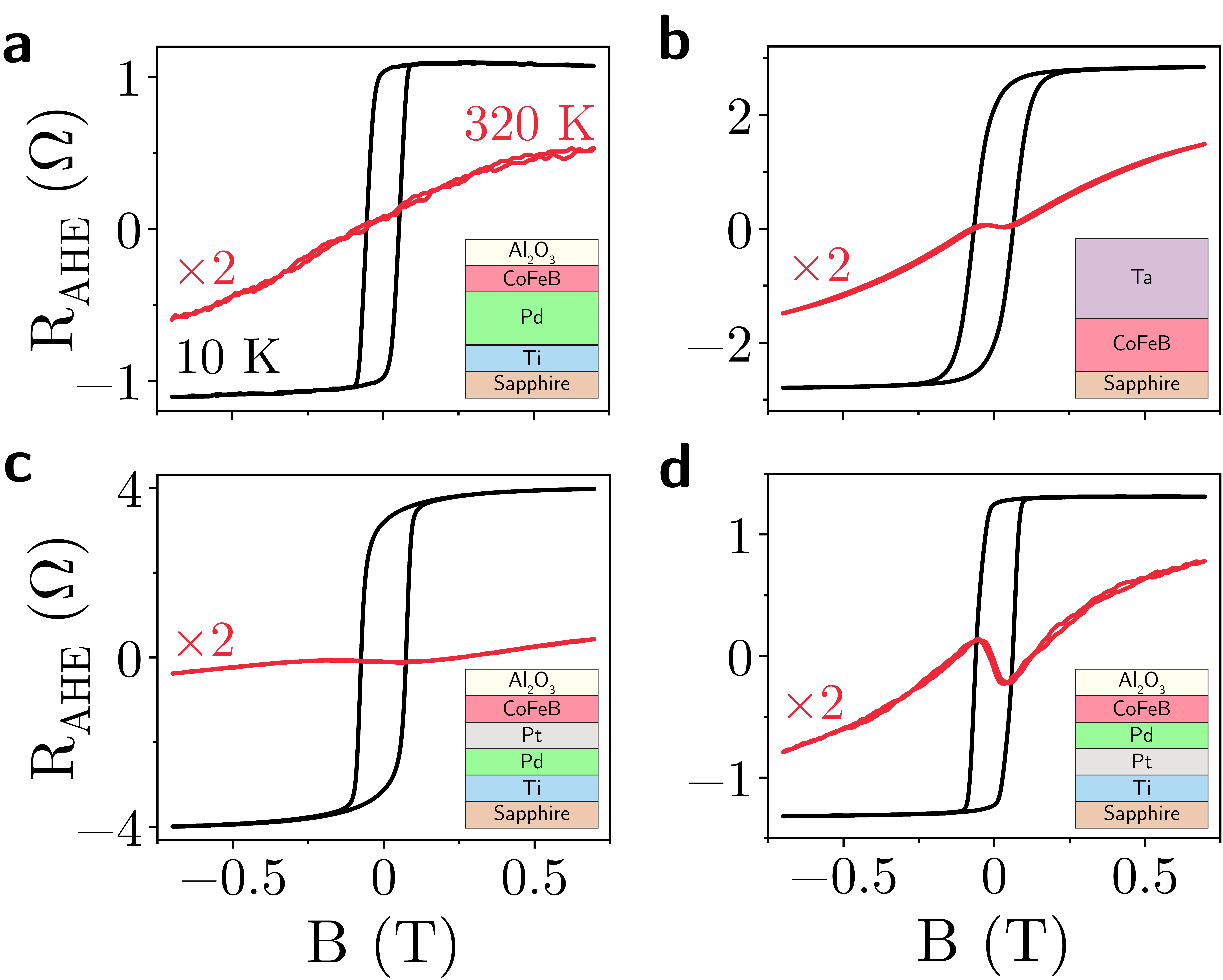}
	\caption{Representative AHE curves for Ti(1.5)Pd(2)/CoFeB(0.3) (a), CoFeB(0.7)/Ta(3) (b), Pd(1)Pt(1)/CoFeB(0.3) (c), and Pt(1)Pd(1)/CoFeB(0.3) (d).}\label{fig:composition}
\end{figure}

Figure~\ref{fig:composition} shows AHE curves for four representative structures elucidating the effects of material properties on the anomalous AHE feature. The feature is not observed when Pt is replaced by Pd, which is isoelectronic with Pt and similarly to Pt exhibits proximity-induced magnetism but much smaller SOC. The central role of SOC is confirmed by the presence of the anomalous feature in a  CoFeB/Ta bilayer [Fig.~\ref{fig:composition}(b)]. Note that the AHE resistance amplitude in Pd/CoFeB and CoFeB/Ta is smaller than in Pt/CoFeB, consistent with a large  contribution to AHE from proximity magnetized Pt, a smaller contribution in Pd due to weaker SOC, and negligible magnetic proximity effects in Ta. The lack of correlation between proximity magnetism and the anomalous AHE feature shows that the latter does  not originate from nonlinear AHE in proximity-magnetized HM~\cite{PhysRevLett.109.107204}. 

The importance of interplay between SOC and magnetism is elucidated by comparing the results for structures Pd(1)Pt(1)/CoFeB and Pt(1)Pd(1)/CoFeB containing the same Pt and Pd layers but in the opposite order. In the FM state, the amplitude of $R_\text{AHE}$ for the former is almost four times larger than for the latter and is similar to Pt(2)/CoFeB [Fig.~\ref{fig:AHE}], confirming a large contribution to AHE from proximity-magnetized Pt within its sub-nm magnetic correlation length~\cite{10.1063/1.4802954}. In contrast, the amplitude of the anomalous feature is larger for Pt/Pd/CoFeB. 
Its enhancement in the structure where Pt is separated from CoFeB by the Pd spacer suggests that it is suppressed by direct exchange coupling of HM to FM. 

\begin{figure}
	\includegraphics[width=0.9\linewidth]{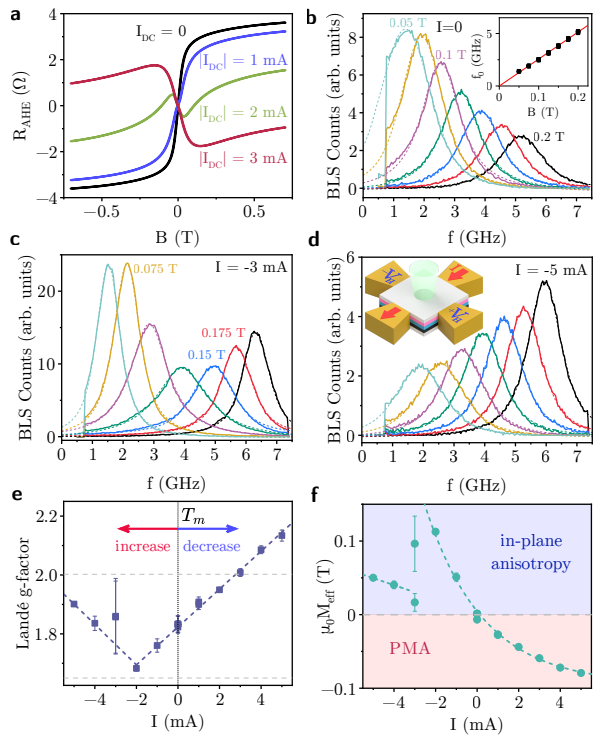}
	\caption{Effect of current in a Ti(1.5)Pt(2)CoFeB(0.4)AlO$_x$(2) film patterned into a $4\mu $m$\times4\mu$m Hall bar, at $T=295$~K. (a) $R_\text{AHE}$ vs. $B$, at labeled $I$. (b)-(d), BLS spectra at labeled in-plane field, at $I=0$ (b), $I=-3\text{mA}$ (c), $I=-5\text{mA}$ (d).  (e),(f) Landé $g$-factor (e) and magnetic anisotropy (f) vs. $I$. Inset in (b): fitting with Kittel formula, and in d: measurement schematic. Dashed lines are guides for the eye. 
BLS measurements were performed on a separate Hall bar identical to that in (a).}\label{fig:BLS}
\end{figure}

 We now present direct evidence for the magnetic origin of the anomalous AHE feature from measurements of the effects of current in a Pt/CoFeB film patterned into a microscale Hall bar. The FM thickness is adjusted so that $T_{c1}$ is close to RT used in these measurements, as confirmed by the nonlinear paramagnetic dependence $R_{\text{AHE}}(B)$ at current $I=0$ with no anomalous contribution [Fig.~\ref{fig:BLS}a]. A sharp nonlinear feature emerges in AHE at finite $I$, rapidly increasing in amplitude with increasing $I$ without significant broadening. This cannot be explained by Joule heating: the feature is sharper and larger in amplitude than is observed without bias at any $T$. 

These results show that current can stabilize a large anomalous contribution not achievable by temperature variations. We investigate its manifestations in magnetic properties by micro-focus Brillouin light spectroscopy ($\mu$-BLS) of magnetic fluctuations [see Methods]. $\mu$-BLS is sensitive to the dynamical component of magnetization normal to the film plane, which was facilitated by measurements with in-plane field transverse to the current [inset in Fig.~\ref{fig:BLS}d]. In this geometry, SOT produced on CoFeB by the Pt layer results in suppression (enhancement) of fluctuations at $I>0$ ($I<0$), which can be approximated as the variation of effective magnetic temperature $T_m$~\cite{PhysRevLett.107.107204}. If the anomalous feature observed in the temperature-dependent AHE curves at $T>T_{c2}$ reflects a true magnetic contribution, one expects to see anomalies in the fluctuation spectra at $I<0$. 

BLS spectra of thermal fluctuations are well-described by the Lorentzian function. Lorentzian fitting was used to determine the central frequency $f_0$ of fluctuations. At $I=0$, $f_0$ increases with field following the usual Kittel formula [see \textit{Methods} and Figure S8], while their intensity decreases as $1/f_0$ in accordance with the Bose–Einstein statistics, Fig.~\ref{fig:BLS}b. In the regime of enhanced $T_{m}$ at $I<0$ the dependence of the detected fluctuation intensity on field becomes highly non-monotonic, with a large dip at a current-dependent field [see Figs.~\ref{fig:BLS}c,d and Supplemental Material]. We interpret this as a manifestation of the anomalous contribution to magnetism, as follows. Since the mechanisms of MOKE and BLS are the same, BLS is not directly sensitive to the anomalous contribution. Consequently, the detected intensity of fluctuations is expected to decrease when this contribution to fluctuations becomes large.

The non-monotonic fluctuation intensity is accompanied by anomalous variations of magnetic anisotropy and Landé $g$-factor, providing evidence for a significant role of SOC which governs both of these characteristics. They were extracted from the Kittel fitting of the dependences $f_0($B$)$ [see Methods]. At $I>-2$~mA, both parameters exhibit smooth variations, which can be attributed to varying relative contributions of CoFeB and proximity magnetized Pt characterized by different anisotropies and $g$-factors~\cite{10.1063/1.2000457,FISCHER19691527}. Both the anisotropy and the Landé $g$-factor  rapidly change at $I<-2$~mA in concert with the suppression of fluctuation intensity attributed to the onset of anomalous magnetic contribution at $T_m>T_{c2}$. The rapid increase of $M_\text{eff}$ reflects a large PMA associated with this contribution, while the variation of $g$-factor suggests a large orbital moment.

\section{\label{sec:Discussion}Discussion}

\begin{figure}
	\includegraphics[width=1\linewidth]{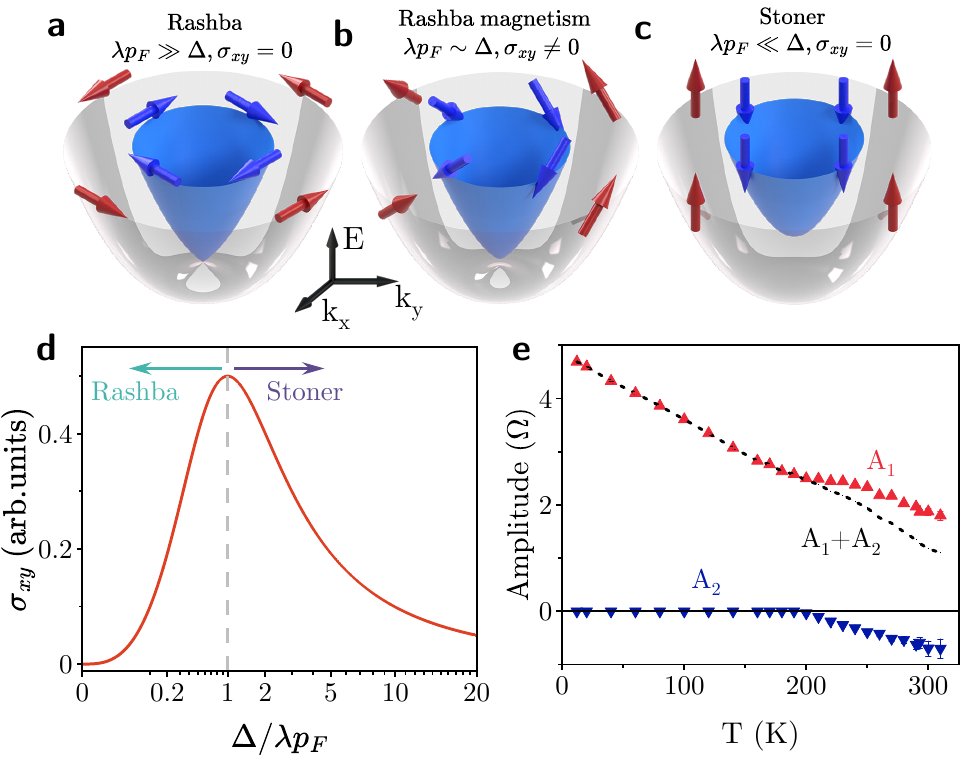}
	\caption{Interplay between Rashba mechanism and magnetism. (a)-(c) Spin texture of Fermi surface in the non-magnetic Rashba state (a), for exchange splitting similar Rashba energy (b), and dominant exchange (c). (d) Dependence of AHE conductivity on the relation between exchange splitting and Rashba energy, (e) Temperature dependence of sigmoid amplitudes for the sample in Fig.~\ref{fig:AHE}.}\label{fig:cones}
 \end{figure}



The existence of anomalous AHE feature in FM/HM bilayers but not in FM/Pd shows that it requires a large SOC. The feature is correlated with ferromagnetism, but does not originate from the proximity magnetized HM volume, as evidenced by the dependence on composition. These observations point to the interplay between SOC and magnetism at the interfaces.

The effects of SOC at interfaces can be described by the Rashba Hamiltonian $H_{R}=\lambda\mathbf{s}\cdot(\mathbf{p}\times\hat{\mathbf{z}})$, where $\lambda$ is the Rashba coefficient, $\mathbf{s}$ and $\mathbf{p}$ are electron spin and momentum, respectively~\cite{Rashba1984}. In the absence of magnetism, a band with effective mass $m$ is split by Rashba SOC into subbands with opposite spin textures, Fig.~\ref{fig:cones}(a). In magnetic systems, normal effective field $\mathbf{B}_\text{eff}$ including external field $\mathbf{B}$ and exchange field $\mathbf{B}_\text{ex}$ cants the spin textures [Fig.~\ref{fig:cones}(b)] and imparts the bands with Berry curvature resulting in a finite Rashba AHE conductivity $\sigma_{xy}=\frac{e^2}{2h}\Delta\left(\frac{1}{\Delta_{p-}}-\frac{1}{\Delta_{p+}}\right)$, where $\Delta_p=\sqrt{\Delta^2+\lambda^2p^2}$, $\Delta=\mu_BB_\text{eff}$ is effective Zeeman energy, and $\Delta_{p\pm}$ is the value of $\Delta_{p}$ for the Fermi momentum $p_F$ of the corresponding sub-band~\cite{RevModPhys.82.1539}. At $\Delta\gg\lambda p_F$, Rashba spin texture becomes suppressed and $\sigma_{xy}$ vanishes, Fig.~\ref{fig:cones}(c).

Using $\Delta,\,\lambda p_F\ll p_F^2/2m$ for the studied metallic films with large Fermi momentum, we obtain  $\sigma_{xy}=\frac{e^2}{h}\frac{\Delta\lambda^2m}{\Delta^2+\lambda^2p^2_F}$. Rashba AHE vanishes in the absence of effective field, reaches a maximum $e^2\lambda m/2h$ at $\Delta=\lambda p_F$, and disappears again at $\Delta\gg\lambda p_F$, Fig.~\ref{fig:cones}(d).

Based on this analysis, we identify the anomalous AHE feature with Rashba AHE at the FM/HM interface. Exchange energy in the FM state of TM FMs is at least an order of magnitude larger than Rashba energy $\lambda p_F$ at the  interfaces of HMs and their compounds, suppressing anomalous contribution~\cite{Schulz2019,BIHLMAYER20063888}. It emerges in the paramagnetic state when the condition $\Delta\approx\lambda p_F$ is satisfied in some field range, resulting in a spin-polarized state characterized by chiral reciprocal-space spin texture which can be described as Rashba magnetism. This state is suppressed at large $\Delta$, consistent with the disappearance of the anomalous AHE feature at $T<T_{c2}$. 

$\Delta\gg \lambda p_F$ can be also achieved by applying a sufficiently large field at $T>T_{c2}$, implying that the anomalous contribution to AHE should decay at large $B$. This is confirmed by analyzing the amplitudes obtained from the two-sigmoid fitting of AHE curves, Fig.~\ref{fig:cones}(e). The amplitude $A_1$ of the dominant contribution extracted from this fitting is correlated with the amplitude $A_2$ of the anomalous contribution, such that their sum $A_1+A_2$ is a smooth almost linear function of $T$. For anomalous contribution that decays at large $B$, $A_1+A_2$ is the true amplitude of the dominant contribution, eliminating this artificial correlation.

Our other unusual observations are also consistent with the proposed mechanism.  MOKE is sensitive predominantly to polarization of quasi-localized d-electrons, while Rashba magnetism is hosted by conduction states, explaining why it shows up only in AHE. Rashba AHE conductivity is independent of the sign of the Rashba coefficient, consistent with the same sign of the anomalous AHE feature in all the studied structures based on the same FM. The decrease of the anomalous AHE amplitude with decreasing $T>T_{c2}$ is explained by the proliferation of magnetic fluctuations with large amplitude $\Delta>\lambda p_F$ suppressing Rashba AHE. The large current-induced Rashba AHE in magnetically ordered state indicates that current suppresses the exchange field experienced by conduction electrons without substantially increasing magnetic fluctuations dominated by the quasi-localized d-electrons. This interpretation warrants further studies of current-induced effects beyond classical magnetization rotation by SOTs~\cite{PhysRevLett.119.257201,PhysRevLett.126.037203}. 

We explain the origin of magnetic transition at $T_{c2}$ in terms of different magnetic anisotropies associated with the two contributions to magnetism. Rashba magnetism is predicted to exhibit a large PMA and orbital magnetization~\cite{Barnes2014}, consistent with the increase of magnetic anistropy and $g$-factor in BLS measurements at $I<-2$~mA [Fig.~\ref{fig:BLS}]. On the other hand, at small $I<0$ corresponding to $T_{c1}<T_m<T_{c2}$ in these measurements, the anisotropy is in-plane. According to the Mermin-Wagner theorem, ordering is suppressed in 2D systems with continuous symmetry of in-plane order parameter~\cite{Kosterlitz1973}. Thus, $T_{c2}$ can be interpreted as an incipient transition which does not result in magnetic ordering because of the loss of PMA associated with Rashba magnetism.

\section{\label{sec:Conclusions}Conclusions}

We have demonstrated an anomalous contribution of Rashba SOC mechanism to magnetism and AHE in ultrathin magnetic film heterostructures. It emerges above the magnetic ordering temperature or under the influence of current, and is suppressed in the magnetically ordered state. The dependence of Rashba AHE on the effective magnetic field provides a tentative explanation for the puzzling observations of nonlinear temperature-dependent AHE features in a variety of magnetic heterostructures~\cite{doi:10.1063/1.2827174,PhysRevB.87.104407,PhysRevLett.109.107204,PhysRevB.89.140407,doi:10.1063/1.4916342,PhysRevB.92.165114,10.1007/s12598-022-02166-z}. The effects identified in our work have significant implications for  SOTs driven by Rashba mechanism. Rashba SOTs are determined by the current-dependent spin polarization associated with Rashba spin texture~\cite{RevModPhys.91.035004}. The latter is suppressed by magnetism, reducing the efficiency of SOTs. It is then possible to control Rashba SOT by the judicious choice of materials reducing exchange splitting at Rashba interfaces. This can be accomplished, for example, by inserting a nonmagnetic spacer between Rashba interfaces and magnetic materials. 

Our study elucidates the long-debated role of magnetism in SOTs~\cite{PhysRevB.98.134406,PhysRevB.97.020403}. Based on our results, Rashba contribution to magnetoelectronic and spin-galvanic effects in typical FM/HM heterostructures is suppressed by magnetism, consistent with some studies~\cite{Tao2018,PhysRevLett.122.077201}. However, this suppression mechanism is affected by exchange splitting in FM and magnetic proximity effects dependent on interface composition and structure, which may explain the apparent contradiction with other evidence pointing to large Rashba effects~\cite{Haidar2019-zc,PhysRevB.104.184410}.

\section{\label{sec:Methods}Methods}

\textbf{Samples.} Ti(1.5)Pt(2)Co$_{40}$Fe$_{40}$B$_{20}$(d)AlO$_x$(2) and other similar heterostructures, where numbers in parentheses are thicknesses in nanometers and $d$ ranges from $0.3$~nm to $0.4$~nm, were fabricated by sputtering in $0.5$~Pa of ultrahigh-purity Ar on sapphire substrates, in an ultrahigh-vacuum chamber with the base pressure of $7\times10^{-7}$~Pa. The Ti buffer layer produced (111) texture of fcc Pt and perpendicular magnetic anisotropy of CoFeB which approximately compensated the demagnetizing field, facilitating precise sigmoid fitting of hysteresis curves without the need for anisotropy corrections. The AlO$_x(2)$ capping layer protected the magnetic layers from oxidation.

\textbf{Measurements.} Magnetoelectronic measurements of AHE in extended films were performed in the van der Pauw geometry on $6$~mm$\times6$~mm square chips, using lock-in detection with ac current of $10~\mu$A. Electronic measurements of the effects of current were performed on films patterned by e-beam lithography into microscale Hall bars with $4~\mu$m$\times4~\mu$m dimensions, using lock-in detection with ac current of $10~\mu$A superimposed with dc current. To separate two contribution to AHE, the curves were fitted with the sum of two sigmoids,
\begin{equation}\label{eq:ahe_fitting}
R_{\text{AHE}}=A_1\tanh\left(B\chi_1\right)+A_2\tanh\left(B\chi_2\right)
\end{equation}
where $A_i$ and $\chi_i$ are amplitudes and susceptibilities (scaled by AHE response) of two contributions, with $i=1,2$. Magneto-optic Kerr effect (MOKE) characterization was performed in a polar configuration using a collimated $2$~mW HeNe laser beam.

Micro-focus BLS measurements were performed at room temperature with an in-plane magnetic field, using the probing laser light with the wavelength of $532$~nm and the power of $0.1$~mW focused into a diffraction-limited spot at the center of the $4~\mu$m$\times4~\mu$m Hall bar by a high-numerical-aperture $100$x microscope objective. The spectrum of light inelastically scattered from magnetization fluctuations was analyzed by a six-pass Fabry–Pérot interferometer TFP-2HC (JRS Scientific Instruments, Switzerland). The obtained BLS intensity at a given frequency is proportional to the square of the amplitude of the normal component of dynamic spin magnetization at this frequency~\cite{4407581}. The dependence of the central frequency of the fluctuation spectra on field was fitted with the Kittel function

\begin{equation}\label{eq:kittel}
f=\frac{g\mu_B}{h}\sqrt{B\left(B+\mu_0M_{\text{eff}}\right)},
\end{equation}
where $h$ is the Planck's constant, to determine the $g$-factor and the net magnetic anisotropy $M_{\text{eff}}$.\newline

The data that support the findings of this study are available from the corresponding author upon reasonable request.

\begin{acknowledgments} S.I., Y.Z, J.P., and S.U. were supported by the NSF Award ECCS-2005786,  sample fabrication and AHE measurements by G.C. were supported by the DOE BES award DE-SC0018976, Y.Z. was supported in part by the SEED award from the Research Corporation and Tarbutton fellowship from Emory College of Arts and Sciences. We thank M. Mourigal for magnetometry. This research used resources of the Advanced Photon Source operated for the DOE Office of Science by Argonne National Laboratory under Contract No. DE-AC02-06CH11357.\newline
\end{acknowledgments}

Author contributions: S.U. conceived the project, sample fabrication and AHE measurements were performed by S.I., G.C. and J.P., magnetometry by M.M., BLS by V.D. and S.D., XMCD by N.B., B.W. and J.W.F., and noise measurements by Y.Z. All authors contributed to the manuscript.\newline

The authors declare no competing interests associated with this research.

\bibliography{Bibliography}
\bibliographystyle{apsrev4-2}
\end{document}